# MODELLING OF GAS SAMPLE BEHAVIOUR IN GAS CHROMATOGRAPHY COLUMN

Z. GNIAZDOWSKI, P. KOWALSKI

INSTITUTE OF ELECTRON TECHNOLOGY, POLAND

**ABSTRACT**: The equilibrium–dispersive model of the linear GC (gas chromatography) was derived using both assumptions and the method of its derivation different from the known in literature. It was concluded that this model is a specific case of the Fokker-Planck equation for diffusion with drift. The resolution of this derived equation for assumed initial conditions is the normal distribution with movable mean value. At the end of the GC column the standard deviation of this distribution was investigated as a factor of the sample shape. This standard deviation is independent of the retention factor.

## INTRODUCTION

Recently gas micro-chromatography systems are under development. Integration of complete system (valves, dosing system, chromatography column, detector) on one silicon chip is the goal. Theoretical analysis of gas sample behaviour in gas chromatography column is the base for this design.

The GC column with carrier gas and the solute is considered. The retention factor $k$ is assumed constant. This factor determines the distribution of the solute between mobile phase and stationary phase.

On the other hand, in the mobile phase the particles of the solute perform the Brownian motions. It is assumed that these motions are in one dimension along capillary. After each motion the quantity of particles are shared between mobile phase and stationary phase in conformity with the retention factor. The problem is description of the solute behaviour. It is necessary to find such a function, which describes the situation of the particle at any moment of time $t > 0$. As the particle performs the random motions, the solution of the problem is the function $u(x,t)$ equal to the probability that at the moment $t$ the particle will be at the point of coordinate $x$. It is the distribution function of the solute in the mobile phase. Additionally, the distribution of the solute in the stationary phase is denoted as $v(x,t)$.

Using these denotations, the retention factor is equal:

$$\frac{v(x,t)}{u(x,t)} = k. \quad (1)$$

From here:

$$v = ku \quad (2)$$

and the quantity of the sample in the mobile phase to the total sample both in stationary and mobile phase is equal:

$$\frac{u}{u+v} = \frac{u}{u+ku} = \frac{1}{1+k}. \quad (3)$$

## THE MODEL

The motion of the particle in the mobile phase is considered. The particle moves in one dimension along the real number axis conjugated with the GC column. For this purpose the reasoning like in [1] and [2], increased by assumptions respected to the retention factor will be performed.

At the moment $t = 0$ the particle is in the origin $x = 0$ of the coordinate system. If in the moment $t$ the particle is situated in the point $x$, then in moment $t + \Delta t$ it is in the point $x + \Delta x$ with the probability $p$, or in point $x - \Delta x$ with probability $q = 1 - p$. In the same way: if the particle in the moment $t$ is situated in the point $x - \Delta x$, then in the moment $t + \Delta t$ the particle will be in the point $x$ with probability $p$. Also, if the particle in the moment $t$ is situated in the point $x + \Delta x$, then in the moment $t + \Delta t$ the particle will be in the point $x$ with probability $q = 1 - p$. The motions of the particles are independent. Therefore, the probability that the particle in the moment $t + \Delta t$ will be situated in the point with coordinate $x$ is equal to the sum of probabilities:

$$u(x, t + \Delta t) = pu(x - \Delta x, t) + qu(x + \Delta x, t). \quad (4)$$

The initial conditions for $t = 0$ are:

$$u(0,0) = 1;\ u(x,0) = 0 \text{ for } x \neq 0 \quad (5)$$

Equation (4) permits to calculate values of $u$ for all $x$ and $t$ with retention factor equal to $0$. The existence of nonzero retention factor affects the modification of the right side of this equation. It is assumed the existence of retention factor in each time and each place in the GC column. Hence, this ratio is obliged also in the while $t + \Delta t$. After the particle motion described above, the total amount of sampled gas in GC column (both in mobile phase and in the stationary phase) have to adapt to condition resulted from existence of retention factor. From here, the density of the gas fulfilled equation (3).

For that reason, it is necessary to modify the right side of equation (4) according to equation (3):

$$u(x, t+\Delta t) = (v(x,t) + u(x, t+\Delta t)) \frac{1}{1+k} \quad (6)$$

The term $v(x,t)$ can be eliminated from (6), using dependence (2). In addition, the term $u(x, t+\Delta t)$ can be replaced by the right side of equation (4). Hence:

$$u(x, t+\Delta t) = \frac{ku(x,t) + pu(x-\Delta x, t) + qu(x+\Delta x, t)}{1+k} \quad (7)$$

Using the terms of Taylor series expansion up to order three, the equation (7) may be written as:

$$u + \frac{\partial u}{\partial t}\Delta t = \frac{1}{1+k}\left\{ \begin{array}{l} ku + \\ pu - p\frac{\partial u}{\partial x}\Delta x + \frac{1}{2}p\frac{\partial^2 u}{\partial x^2}(\Delta x)^2 \\ + qu + q\frac{\partial u}{\partial x}\Delta x + \frac{1}{2}q\frac{\partial^2 u}{\partial x^2}(\Delta x)^2 \end{array} \right\} \quad (8)$$

Dividing both sites of this equation by $\Delta t$ and doing simplification:

$$\frac{\partial u}{\partial t} = \frac{1}{1+k}\left\{ -(p-q)\frac{\partial u}{\partial x}\frac{\Delta x}{\Delta t} + \frac{1}{2}\frac{\partial^2 u}{\partial x^2}\frac{(\Delta x)^2}{\Delta t} \right\} \quad (9)$$

Equation (9) may be transformed by passing to the border $\Delta t \to 0$ in order to [1] and [2]:

$$D = \frac{1}{2}\frac{(\Delta x)^2}{\Delta t}, \quad c = (p-q)\frac{\Delta x}{\Delta t} \quad (10)$$

In this way:

$$\frac{\partial u}{\partial t} = \frac{1}{1+k}\left\{ -c\frac{\partial u}{\partial x} + D\frac{\partial^2 u}{\partial x^2} \right\} \quad (11)$$

Denoting $D' = D/(1+k)$ and $c' = c/(1+k)$ equation (11) has a form:

$$\frac{\partial u}{\partial t} = -c'\frac{\partial u}{\partial x} + D'\frac{\partial^2 u}{\partial x^2}. \quad (12)$$

## DISCUSSION

The analogous formula derived using the mass balance method is known in literature [3], [4]. Formula (12) is the specific case of Fokker-Planck equation for diffusion with drift [1], [2], [5], where $c'$ is drift velocity and $D'$ is diffusion coefficient of the sample in the carrier gas. For $c = 0$ the roaming of the particles is symmetrical. The sign of number $c$ defines the direction of the drift.

Effective drift velocity of this sample is $(1+k)$ times less than velocity of the carrier gas. It means: for assumed constant $k$ the retention time $t_R$ (it is an average time of the stay of gas sample in the column) is $(k+1)$ times greater than $t_0 = L/c$ - the time of the stay of un-retained gas (also carrier gas) in the column. Hence, $k = t_R/t_0 - 1$.

The resolution of equation (12) for conditions (5) is the probability density function $u(x,t)$:

$$u(x,t) = \frac{1}{\sqrt{4\pi D't}}\exp\left(-\frac{(x-c't)^2}{4D't}\right) \quad (13)$$

It is the normal distribution with the standard deviation $\sigma = \sqrt{2D't}$ and movable mean value $m = c't$.

The sample with retention factor $k$ and diffusion coefficient of the sample in carrier gas equal to $D$ is considered. At the end of the GC column the standard deviation of the sample is:

$$\sigma = \sqrt{2D't_R} = \sqrt{2\frac{(1+k)Dt_0}{1+k}} = \sqrt{2Dt_0} \quad (14)$$

This standard deviation is independent of the retention factor. It means: for two different samples with the same diffusion coefficients and different retention factors, the shapes of the samples at the end of the column will be the same.

## CONCLUSIONS

For a better understanding and for future optimisation of GC column and detection system performances, the equilibrium–dispersive model of linear GC was investigated. For assumed initial conditions the model of the gas sample distribution was derived and its shape of the end of the GC column was considered. It was concluded that this shape is independent of the retention factor. Based on above investigations, the future work will be focused on the investigation of the behaviour of different gas samples in different sampling conditions, in the GC column.

## THE AUTHORS

Dr. Zenon Gniazdowski and Mr. Paweł Kowalski are with the Institute of Electron Technology, Al. Lotnikow 32/46, 02-668 Warsaw, Poland.
E-mail: gniazd@ite.waw.pl

## REFERENCES

[1] W. Feller: An introduction to Probability Theory and Its Application, Vol.1, J. Wiley and Sons, Inc. 1961
[2] R. Zieliński: Metody Monte Carlo, PWN 1970
[3] A. Felinger: Data Analysis and Signal Processing in Chromatography. Elsevier 1997
[4] H. Gunzler., A. Williams., ed.: Handbook of Analytical Techniques, WILLEY-VCH 2001
[5] H. Risken: The Fokker Planck Equation. Methods of Solution and Applications, Springer 1996